\documentclass[]{raa}

\usepackage{array}
\usepackage{booktabs}
\usepackage{graphicx, times}
\usepackage{natbib}
\usepackage{amssymb,amsmath}
\usepackage{bm}
\usepackage{xcolor}
\usepackage{siunitx}
\usepackage{subcaption}

\newcommand\rjlr[1]{ \textcolor{red} {{\bf #1}} } 

\newcommand{\atlas}{ATLAS$^{\rm 3D}$}
\begin{document}

\title{Constraining Schwarzschild Models with Orbit Classifications}


\author{Richard J. Long}


\institute{Department of Astronomy, Tsinghua University, Beijing 100084, China; {\it rjlastro@yahoo.com; Orcid:0000-0002-8559-0067} \\
           }

\date{Received~~2024 month day; accepted~~2024~~month day}

\renewcommand{\labelitemi}{$\bullet$}
\newcolumntype{M}[1]{>{\centering\arraybackslash}m{#1}}

\abstract{
A simple orbit classification constraint extension to stellar dynamical modeling using Schwarzschild's method is demonstrated.
The classification scheme used is the existing `orbit circularity' scheme ($\lambda_z$) where orbits are split into four 
groups---hot, warm, cold and counter-rotating orbits.  Other schemes which can be related to the orbit weights are expected to be viable as well.  The results show that the classification constraint works well in modeling.  However, given that orbits in external galaxies are not observable, it is not clear how the orbit classification for any particular galaxy may be determined.  Perhaps range constraints for different types of galaxies determined from cosmological simulations may offer a way forward.
}

\authorrunning{Richard J. Long}            
\titlerunning{Orbit Classifications and Schwarzschild Modeling}  

\maketitle


\section{Introduction}
\label{sec:intro}

The stellar dynamics of galaxies may be investigated by creating models of galaxies, constraining them with observational data, and then examining the models to see what might be learnt about the real galaxies.  The book \cite{BT2008} describes the theory behind some of the traditional modeling techniques that might be used. A key point which must be accepted about external galaxy models is that the models can only be indicative or illustrative of the galaxies because of the current technical limitations of the instruments employed for observations: six-dimensional (6D) galaxy phase space data are not obtainable. Mathematically, recovery of the phase space data from observations is not possible in general.  Related deprojection issues are included, for example, in \cite{Rybicki1987}, and, more recently, in papers by \cite{Cappellari2020} and \cite{Vasiliev2020}.  \cite{Long2021} demonstrated that Schwarzschild's method does not comply fully with the theoretical statements in \cite{Rybicki1987}.  Observing and modeling our Galaxy are different in that the instrumental capability to obtain three-dimensional (3D) spatial and velocity data is available. 
As a consequence, for example, it is not clear to what extent orbits in galaxies in cosmological simulations are truly representative of those present in real galaxies.
Specialized n-body simulations help us gain insight into the orbits that might comprise particular components of galaxies, e.g. galactic bars, but do not give us the capability to model all aspects of real galaxies accurately.

It has been recognized that applying machine learning modeling methods to physical systems requires that the relevant physical laws should be adhered, and has resulted in, for example, the development of physics informed neural networks \citep{Raissi2019}.  It is appropriate to revisit more traditional stellar dynamical modeling methods, such as Schwarzschild's method \citep{Schwarz1979}  or Syer \& Tremaine's made-to-measure method \citep{Syer1996}, to understand whether their solutions are astrophysically realistic or not.  In this investigation, Schwarzschild's method which superimposes weighted stellar orbits to try and reproduce observational data is examined.

Using Schwarzschild's method to estimate the global properties of galaxies (e.g. total mass) is well-developed but working with the weighted orbits comprising the Schwarzschild solution, perhaps to help understand the evolutional history of a galaxy, does need careful consideration \citep{Jin2019, Jin2020}. The process used to determine the orbit weights is dealing with an under-determined optimization problem: there are few observational constraints by comparison with the number of orbits. Also, it must be remembered that Schwarzschild's method was created to demonstrate that some self-consistent models of triaxial galaxies could be created: no claims were made on how astrophysically realistic such models would be \citep{Schwarz1979, BT2008}.

Operationally, the known shortcomings of Schwarzschild's method include the following.
\begin{itemize}
\item Low orbit utilization (without regularization) in that only $\approx 10\%$ of orbits supplied are an effective part of any solution---see Table 2 of \cite{Jin2019}.  This percentage becomes even lower if chemo-dynamical techniques such as in \cite{Long2018} are applied.  Regularization can be used to improve the utilization to typically $\approx 75\%$ or more depending on the optimization method.
\item The orbit weighting scheme is purely a numerical process with the resulting weights showing a large weight imbalance in that a small number of orbits receive high weights and conversely many orbits have low or zero weights \citep{Long2018}.
\item Significant overfitting of galaxy data with mean $\chi ^2$ values of $10^{-1}$ or less is not uncommon.
\item Difficulties in determining confidence intervals on model parameters \citep[][section 4.7.2]{BT2008}.
\end{itemize}

In short, the orbit weights have numerical importance to the optimization process (weights act to scale individual orbit contributions to model observables) but whether the orbits individually or collectively as a set are realistic physically and truly exhibit the orbit structure of the galaxy being modeled remains unclear. There are no `necessary and sufficient' statements on representation in the literature regarding the orbits selected.  The weighted orbits can only reflect what is required by the optimization's objective function: if the function is concerned with reproducing observables then the orbits are important in that context but that is all.  As noted in \cite{Jin2019}, for example, they are illustrative or indicative of what a galaxy might be like but cannot be taken as definitive or complete. By extrapolation, any orbit classification applied to the galaxy has to be caveated in the same way.

Given that Schwarzschild's method cannot be used without amendment or enhancement to determine the overall orbit classification for a galaxy, if the orbit classification were to be known from other sources, could that classification be included in Schwarzschild modeling as a constraint perhaps?  Answering this question is the focus of this article.
The article's structure is as follows.  Section \ref{sec:approach} describes at a top level the approach taken in the investigations.   The galaxy data used are described in Section \ref{sec:galdata}, and the relevant theory and descriptions of the methods are in Section \ref{sec:methods}. Results and subsequent discussion are in Sections \ref{sec:cresults} and \ref{sec:discussion}, with conclusions in Section \ref{sec:conclusions}.

\section{Approach} 
\label{sec:approach}

Whereas the Introduction, Section \ref{sec:intro}, sets the context and describes \textit{what} will be investigated, this section describes \textit{how} it will be done.
A number of axisymmetric models of galaxies are constructed, under varying conditions, using an existing software implementation of Schwarzschild's method taken from \cite{Long2021}, and analyzed.  Line-of-sight luminosity and kinematical data from nine galaxies in total are used, with four real galaxies from \atlas\ \citep{AtlasI} and five simulated galaxies from IllustrisTNG \citep{Nelson2019}.  The simulated galaxies have known orbit circularity classifications \citep{Xu2019}.

The initial conditions (spatial and velocity) for the orbits are constructed as in \cite{Long2021}. The three integral (3I) scheme is used with simple adjustments to reflect the degree of bulk rotation of the galaxies, or to vary the number of circular orbits. 
Orbit weights are determined by employing Lawson \& Hanson's non-negative least squares method which uses an active set algorithm internally (NNLS).  The objective function is extended to include not only the usual reproduction of observational data but the reproduction of the target orbit classification as well.  Other optimization schemes could have been used but the NNLS scheme (without regularization) is computationally fast enough to be used in a grid search for orbit classification values.

Modeling runs are accomplished in three stages:
\begin{enumerate}
\item orbit integration and the collection of orbit contributions to the model observables,
\item weight determination using the NNLS optimization method, and
\item analysis of the modeling run, including the calculation of observable $\chi ^2$ values and examining the degree of conformance to the target orbit classification.
\end{enumerate}

In view of the under-determined nature of the weight determination, regularization may be used to restrict the weights' solution space. Here, some models are run both with and without regularization so the impact of regularization is understood. Regularization takes two forms.  The first penalizes heavy weights, and acts to increase the number of orbits actively contributing to the objective function.  The second smooths the particle weights in integral space by penalizing high second derivative values.  These two forms of regularization can be used separately or combined to give a composite form.

There are aspects of stellar dynamical modeling with Schwarzschild's method that are out of scope for this investigation, for example chemo-dynamical modeling \citep{Long2018} or the use of discrete data \citep{Chaname2008}.  Note that only axisymmetric (oblate) galaxy models are considered: the orbit weighting numerical optimization processes are not affected by choice of gravitational potential.  Also, orbit classifications in this work are not targeted at addressing any issues to do with self-consistency, or data deprojection non-uniqueness.

\section{Galaxy Data}
\label{sec:galdata}

For the \atlas\ galaxies (NGC 1248, NGC 3838, NGC 4452, and NGC 4551), the data have been produced by the \atlas\ survey \citep{AtlasI}. The four galaxies are the same galaxies as used in \cite{Long2016} and \cite{Long2018}, and have various inclinations, different mass-to-light ratios, and sense of bulk rotation (see Table \ref{tab:atlasdata}).  Surface brightness values are calculated, using multi-Gaussian expansions (MGEs) (Sect. \ref{sec:gravpot}), on a $(16, 16)$ polar grid with equal interval radial distances, and no straddling of a galaxy's major axis.  The integral field unit (IFU) cells for kinematical data (first and second velocity moments) are the Voronoi cells resulting from signal-to-noise processing of the raw data.
 
\begin{table}
	\centering
	\caption{Top Level Data for the \atlas\ Galaxies}
	\label{tab:atlasdata}
	\begin{tabular}{ccccccc}
		\hline
		Galaxy & Inclination & M/L Ratio & Type & Bulk Rotation & SB Max Radius & IFU Cells\\
		\hline
		NGC 1248 & $42 ^{\circ}$ & $2.50$ & S0 &  Counter clockwise & 3.00 & 297 (1.12, 0.98)\\
		NGC 3838 & $79 ^{\circ}$ & $4.00$ & S0 &  Clockwise & 5.86 & 383 (1.70, 1.09)\\
		NGC 4452 & $88 ^{\circ}$ & $5.20$ & S0 &  Clockwise & 5.17 & 489 (1.09, 0.66)\\
		NGC 4551 & $63 ^{\circ}$ & $4.89$ & E  &  Counter clockwise & 3.41 & 596 (1.27, 1.10)\\
		\hline
	\end{tabular}
	
\medskip
The inclinations are taken from \citet{AtlasXV}, while the mass-to-light ratios (in the M/L Ratio column) were determined by made-to-measure modeling in \cite{Long2016}.  The SB maximum radius column gives the overall size of the polar grid used for surface brightness data values.  The IFU column gives the number of Voronoi cells for the kinematical data.  The numbers in brackets are the maximum `on sky' X and Y values from the galaxy's cell centroids, and give a rough indication of the kinematical data coverage without needing to refer to the convex hull surrounding the cell centroids.
\end{table}

The same IllustrisTNG galaxies \citep{Nelson2019} as in \cite{Long2021} are used.  The selection strategy for the galaxies is described in \cite{Long2021}, and is not repeated here.  The five axisymmetric galaxies (with identifiers, A490, A1090, A1190, A1290, and A1390) are viewed edge on with the mass-to-light ratio taken as $1.0$ for all galaxies.  It has been assumed that the kinematics of stellar particles can validly be used to create stellar kinematical data as if from a real galaxy (consistent with \cite{Li2016} or \cite{Xu2019}, for example).  Surface brightness values are calculated, using MGEs, on a $(16, 16)$ polar grid with logarithmic interval radial distances, and straddling the galaxy's major axis.  The IFU cells for kinematical data (the Gauss-Hermite coefficients $h_1$ to $h_4$) are constructed to mimic a real galaxy's data. Table \ref{tab:simdata} is the equivalent of Table \ref{tab:atlasdata} for the simulated galaxies.
 
\begin{table}
	\centering
	\caption{Top Level Data for the Simulated Galaxies}
	\label{tab:simdata}
	\begin{tabular}{ccccccc}
		\hline
		Galaxy & Bulk Rotation & SB Max Radius & IFU Cells\\
		\hline
		A490 &  Counter clockwise & 7.00 & 220 (2.77, 1.38)\\
		A1090 & Counter clockwise & 6.00 & 197 (2.74, 1.63)\\
		A1190 & Clockwise         & 6.00 & 110 (2.50, 1.71)\\
		A1290 & Clockwise         & 6.00 & 173 (2.77, 1.35)\\
		A1390 & Clockwise         & 7.00 & 119 (2.54, 1.41)\\
		\hline
	\end{tabular}
	
\medskip
As in Table \ref{tab:atlasdata}, the SB maximum radius column gives the overall size of the surface brightness polar grid, and the IFU column gives the number of Voronoi cells with   the numbers in brackets being the maximum X and Y values from the galaxy's cell centroids.  Inclinations for all the simulated galaxies are $90 ^{\circ}$, and mass-to-light ratios are $1.0$.
\end{table}

The IFU cells for the \atlas\ galaxies and the IllustrisTNG simulated galaxies contain the observational kinematical data that Schwarzschild's method is asked to replicate.
Data are symmetrized appropriately for an axisymmetric model/potential.  The design matrix size used in the convex optimization (see equation \ref{eqn:orbwts}) may be calculated as the total number of data points for observables times the number of orbits.  As an example, NGC 3838 has surface brightness (256 points) and two IFU observables (2 x 383 = 766 points) giving 1022 data points in total.  For 8000 orbits, the design matrix size is 1022 rows and 8000 columns.  For simulated galaxy A1190 with four IFU observables, the matrix size is 696 rows and 8000 columns.

\section{Methods} 
\label{sec:methods}

\subsection{Gravitational Potentials}
\label{sec:gravpot}

Galaxy gravitational potentials and derivatives are derived from MGEs of the galaxies' surface brightness \citep{Emsellem1994, Cappellari2002}.  For computer performance reasons, an MGE potential and its gradients are implemented as a set of interpolation tables, for example as in  \cite{Remco2008} or \cite{Long2012}.  

For the simulated galaxies, dark matter potentials (calculated from the simulation's dark matter particles) are included as spherical Gaussians. No black hole modeling takes place.  For the \atlas\  galaxies, no dark matter is included (for consistency with \cite{AtlasXV}), and central black holes are modeled as point sources.

\subsection{Orbit Initial Conditions}
\label{sec:ics}

The initial phase space coordinates for orbits are created using the 3I scheme.  Three variants are created by varying the sense of rotation of the orbits:  
\begin{enumerate}
\item no overall sense of rotation about the symmetry axis, that is 50\% of orbits with one sense of rotation and 50\% with the other;
\item with the sense of rotation matching that shown by a galaxy's mean line-of-sight velocity data for 75\% of the orbits with 25\% counter-rotating; and
\item as in (2) but with 100\% matching the data's sense of rotation and no counter-rotating orbits.
\end{enumerate}

Similarly to \cite{Zhu2018}, the circularity measure $\lambda _z$, the ratio of the orbit's $L_z$ component of angular momentum to the maximum value allowed by the orbit's energy, is used to classify orbits into one of hot ($\left|\lambda _z\right| <= 0.25$), cold ($\lambda _ z >= 0.80) $, warm ($0.25 < \lambda _z < 0.8$) or counter-rotating (the remainder).  Practically, internally, it is more convenient to split the orbits into six classes---rotating or counter-rotating and then one of hot, warm or cold---and only combine classes where necessary for presentational or comparison purposes.  These classes are used extensively in analyzing modeling runs. The total orbit weights in each of the classes are taken as the overall orbit classification for a galaxy.

Note that the combination of a steady state gravitational potential for a galaxy and the initial spatial and velocity conditions for an orbit means that the orbit's trajectory is fully determined. From Hamiltonian mechanics, energy is a conserved quantity for an orbit, and, in addition, for an axisymmetric galaxy potential so is the component of angular momentum ($L_z$) parallel to the symmetry axis of the galaxy. This means that the $\lambda _z$ classifier of orbits as hot, warm, or cold can be found directly from the initial conditions.  As a consequence, if all the orbits in a model are regarded as active orbits (not zero-weighted) by the optimization method then the relative proportions of the numbers of hot, warm and cold orbits must match the proportions from the initial conditions. This remains true even for gravitational systems where $L_z$ is not conserved but some time or orbit averaged value is used instead.  What is of more interest is the relative proportions of the orbit weights allocated to hot, warm and cold orbits as these proportions can be related to the global properties of a galaxy.

\subsection{Schwarzschild Models}

As previously stated, the implementation of Schwarzschild's method is as in \cite{Long2021} and the details are not repeated here.  For simulated galaxies, the observational data utilized are surface brightness with kinematics (Gauss-Hermite coefficients) as in \cite{Rix1997} and \cite{Cretton1999}.  For the \atlas\ galaxies, modeling is with surface brightness and the first and second velocity moments.  All models use $8000$ orbits. No use is made of orbit dithering or point symmetries in collecting orbit contributions to model observables.  Note that observational data are not available to be used to constrain individual orbits.  If they were, they could be included in the objective function to be optimized.  Mixes of orbits by type distribution can be included, as described below in Section \ref{sec:orbittypes}, provided such techniques are linear in superposition in the same way that other observables are. This is believed to be a new extension to Schwarzschild's method, and so has its own section.

Orbit weights are determined using NNLS.  Taking equation 1 in \cite{Long2018}, in matrix form, the expression to be minimized is
\begin{equation}
	\|\mathbf{Dw} - \mathbf{K}\|^2_2,
	\label{eqn:orbwts}
\end{equation}
where $\mathbf{D}$ is the `design' matrix giving the orbit contributions to the model observables, $\mathbf{w}$ represents the orbit weights to be determined, and $\mathbf{K}$ contains the `measured' or target observables.  The L2-norm (Euclidean norm) is indicated by $\|.....\|_2$. 

Note that the Schwarzschild's method can only weight the orbits it has been provided with.  In that respect, the gravitational potential and the orbit initial conditions play a substantial role in determining the eventual solution from Schwarzschild's method.

\subsection{Constraints using Orbit Classifications}
\label{sec:orbittypes}

In this section, the weighting process is first considered in optimization objective function terms (rather than matrices). Usually the Schwarzschild objective function is framed in terms of reproducing the observed luminosity and kinematical data so the optimizing process selects and weights orbits to try and cause this reproduction to happen.  If the objective function is framed to achieve a different end effect (for example, related to the distribution of some other quantity), a set of orbits appropriate to that objective would be selected and weighted.  What is needed here is that not only are observables reproduced but that some astrophysically realistic mix of orbit types is used in so doing.  

The circularity measure $\lambda_z$ can be used to define orbit types and is a convenient place to start.  Mathematically, an expression to be optimized might be  
\begin{equation}
\label{eqn:extcirc1}
\frac{1}{2}\sum_j(\sum_i (w_i \delta(i \in C_j)) - T_j)^2, 
\end{equation}
where $C_j$ is an orbit type within the classification scheme being used, and $T_j$ is the target weight to be met by the orbits in $C_j$.  The selection function $\delta(i \in C_j)$ takes the value $1$ if orbit $i$ has type $C_j$ and $0$ otherwise.  If there are uncertainties $\sigma_j$ associated with the $T_j$ then this can be accommodated simply as in expression \ref{eqn:extcirc}.
\begin{equation}
\label{eqn:extcirc}
	\frac{1}{2} \sum_j \left( \frac{ \sum_i w_i \delta(i \in C_j) - T_j}{\sigma_j} \right)^2.
\end{equation}
Structurally, expression (\ref{eqn:extcirc}) has the same $\chi ^2$ form as the observable data in the objective function.  In this article, expression (\ref{eqn:extcirc1}) is used, together with an orbit classification parameter 
$\mu_{\rm{oc}}$, to set the overall importance of the expression amongst all the other terms in the function.  Here, $\mu_{\rm{oc}} = 10^3$.

Reverting now to matrix terms, each class of orbit type $C_j$ is represented in matrix $\mathbf{D}$ (see equation \ref{eqn:orbwts}) by a row with each column entry having value $1$ if orbit $i$ is in $C_j$ and $0$ otherwise.  The corresponding entries in matrix 
$\mathbf{K}$ are the $T_j$ values.  In the context of $\lambda_z$ circularity classifications, the [red, warm, cold, counter-rotating] target values might, for example, be [30\%, 45\%, 15\%, 10\%] with four matrix rows being required, and the entries per row indicating whether the individual orbits are of the associated type (entry value = 1) or not (entry value = 0).  Note that the mechanism described is not specific to $\lambda_z$ classifications but could be used with other classification schemes as well.
Also, the approach described here is readily adaptable to the made-to-measure method.

\subsection{Regularization}

Two forms of regularization constraint \citep{Tikhonov1963} are employed, penalizing heavy weights and smoothing the weight distribution.  Penalizing heavy weights is as described in \cite{Long2018} and has two effects. It acts to reduce the maximum orbit weight, and as a consequence causes more orbits to have non-zero weights.  Including this regularization term in equation \ref{eqn:orbwts} gives
\begin{equation}
	\|\mathbf{Dw} - \mathbf{K}\|^2_2 + \mu_{\rm{p}} \|\mathbf{w}\|^2_2,
\end{equation}
where $\mu_{\rm{p}}$ is a positive parameter controlling the amount of regularization. As a  constraint, the term is equivalent to $w_i = 0 \; \forall i$. Extending the regularization term to include priors $W_i$ on the orbit weights changes the constraint term to $w_i/W_i = 1 \; \forall i$.  Such regularization has been used elsewhere, for example, in \cite{Valluri2004} and \cite{Vasiliev2013}.  

Smoothing the weight distribution is performed in integral space and is described below.  For axisymmetric models, because of uncertainties about quite what form the third integral takes, only the conserved quantities energy ($E$) and the $L_z$ component of angular momentum are used. Working with the initial conditions on the orbits, each orbit's nearest two neighbors in $(E, L_z)$ space are determined, and the second derivatives of the weights using finite difference formulae are calculated at the orbit's $(E, L_z)$ position.  A two-dimensional tree is used to find the nearest neighbors and the finite difference formula used (backward, forward or central) depends on the relative positioning of the orbit and its neighbors.  Solution or curvature smoothing is a well-established technique, see \cite{CVX} or \cite{NumRecipes2007}, for example.

The regularization parameters are $1.6\times10^3$ for penalization and $6.4\times10^2$ for smoothing, and were determined by experimentation.  For convenience, the same values are used for all galaxies.  Note that both forms of regularization can be used individually in isolation or combined together to 
give a composite regularization term.  For the remainder of this article, the terms \textit{regularization} (for heavy weights penalization) and \textit{smoothing} (for second derivative smoothing) are employed.

\section{Results}
\label{sec:cresults}

\subsection{Baseline Models}
\label{sec:cbaseline}

A set of models is constructed and analyzed to establish a baseline of understanding, or a reference point, for subsequent modeling.  Both sets of galaxies are used for this task.  3I initial conditions, with the rotation of 75\% of the orbits aligned to the bulk rotation of the galaxies, are employed. The orbit classification resulting from the weighted orbits is calculated and, for the simulated galaxies, compared with the classification in \cite{Xu2019}.

\begin{figure}[h]
    \centering
    \caption{Weight progression for galaxy A1190}
	\label{fig:kins1centre}
    \begin{tabular}{ccc}
        \includegraphics[width=70mm]{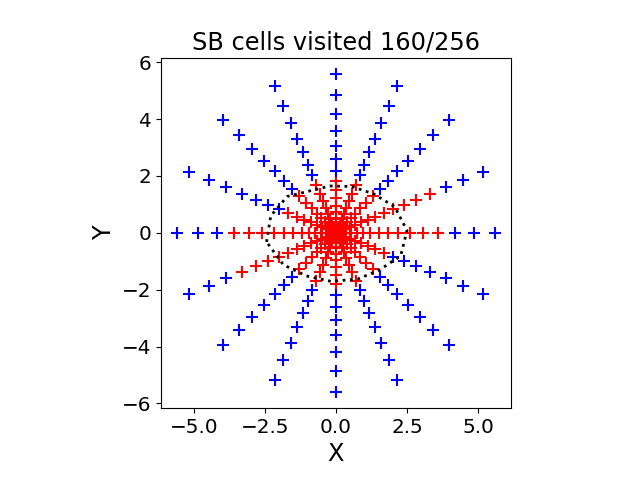}   & \includegraphics[width=70mm]{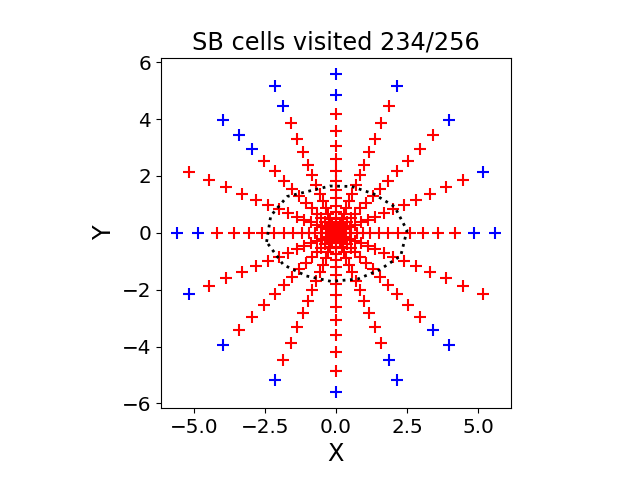} \\
    \end{tabular}
    
\medskip
The plots show the polar grid for the surface brightness observable.  Red indicates that the related grid cell has been visited by one of more orbits; blue indicates that the cell has not yet been visited by an orbit.  The black dashed polygon is the convex hull for the kinematical (IFU)  observables.  
From left to right, the plots show the cells visited by the heaviest orbits comprising 25\% and 50\% of the total weight.  The 75\% and 100\% plots are not included here as all cells have been visited. See Section \ref{sec:cbaseline} for more information.
\end{figure}

\begin{figure}[h]
    \centering
    \caption{Circularity orbit classification for galaxy A1390}
	\label{fig:kins1A1390}
    \begin{tabular}{cc}
        \includegraphics[width=70mm]{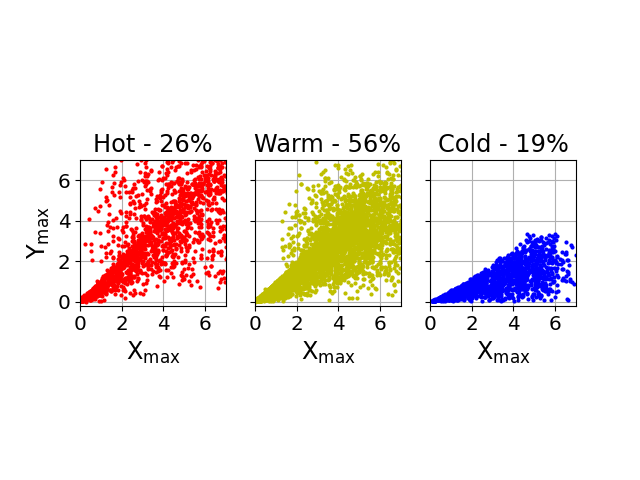}  & \includegraphics[width=70mm]{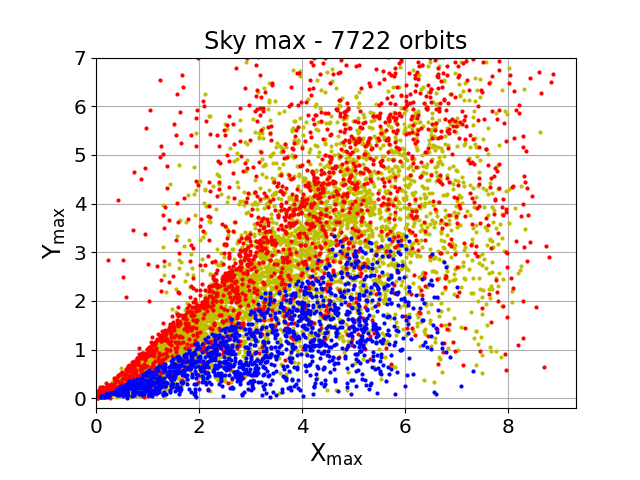} \\
    \end{tabular}

\medskip
Red shows hot orbits, yellow warm orbits, and blue cold orbits.  $(X_{\rm{max}}, Y_{\rm{max}})$ are `on sky' spatial coordinates giving the maximum $X$ and $Y$ positions reached by an orbit.  These coordinates can be used to configure a rectangle which gives a projected region in which the orbit can be found.  The left hand panel gives a separation by orbit class of the right hand panel.  What is apparent from the plots is that cold orbits have a role to play in occupying higher $X_{\rm{max}}$, lower $Y_{\rm{max}}$ regions while hot orbits have a similar role for lower $X_{\rm{max}}$, higher $Y_{\rm{max}}$ regions. See Section \ref{sec:cbaseline} for more information.
\end{figure}

The results for the simulated galaxies are shown in Table \ref{tab:simgalres}.
\begin{table}
	\centering
	\caption{Results for Simulated Galaxies}
	\label{tab:simgalres}
	\begin{tabular}{cccccccc}
		\hline
		Galaxy & & Hot & Warm & Cold & Counter  & Active & Mean\\
		       & &     &      &      & Rotating & Orbits & $\chi^2$\\
		\hline
		A490 &  \\					
		& Initial conditions & 0.25 &	0.41 &	0.16 &  	0.19 &	8000  \\	
		& \cite{Xu2019} & 0.35 &	0.28 &	0.14 &  	0.23  \\						
		& Without reg & \rjlr{0.38} &	0.28 &	0.16 &	\rjlr{0.18} &	743 &	0.071  \\					
		& With reg & 0.33	 &  	\rjlr{0.33}	 &	0.15 & 	\rjlr{0.19} &	4720 & 	0.635  \\
		& With constraint & 0.35 &	0.28 &	0.14	 &  0.23 & 752 & 0.074  \\
		A1090 & \\			
		& Initial conditions & 0.24 &	0.42 &	0.15 &	0.19 &	8000  \\	
		& \cite{Xu2019} & 0.28 &	0.46 &	0.15 & 	0.10	  \\							
		& Without reg & 0.26	 & 	\rjlr{0.39} & 	\rjlr{0.21} & 	\rjlr{0.14} &	658 &	0.064  \\				
		& With reg & 0.28 & 	\rjlr{0.41} & 	\rjlr{0.19} &	0.12 &	5313 & 	0.352  \\
		& With constraint & 0.28 &	0.46 &	0.15 &	0.10 & 651 & 0.070  \\		
		A1190 & \\					
		& Initial conditions & 0.26 &	0.41 &	0.15 &	0.19 &	8000	  \\	
		& \cite{Xu2019} & 0.33 &	0.43 &	0.11 &	0.12  \\					
		& Without reg & \rjlr{0.39} &	\rjlr{0.34} &	\rjlr{0.14} &	0.13 &	559 &	0.015  \\					
		& With reg & \rjlr{0.37} &	\rjlr{0.38} &	0.11 &	0.14 &	5843 &	0.374  \\
		& With constraint & 0.33	 &	0.43	 &	0.11 &	0.12 & 548 & 0.017  \\		
		A1290 & \\					
		& Initial conditions & 0.26 &	0.41 &	0.15	 &	0.19 &	8000  \\	
		& \cite{Xu2019} & 0.32 &	0.37	 &	0.18 &	0.13  \\						
		& Without reg & \rjlr{0.27} &	\rjlr{0.33} &	\rjlr{0.26} &	0.14 &	671 &	0.030  \\				
		& With reg & \rjlr{0.27} &	0.37 &	\rjlr{0.22} &	0.15 &	5030 &	0.328  \\
		& With constraint & 0.32 &	0.37 &	0.18 &	0.13 & 667 & 0.036  \\	
		A1390 & \\				
		& Initial conditions & 0.25 &	0.42 &	0.14 &	0.19 &	8000  \\	
		& \cite{Xu2019} & 0.29 &	0.45 &	0.15 &	0.12  \\						
		& Without reg & 0.27 &	\rjlr{0.41} &	\rjlr{0.18} &	0.14 &	529 &	0.039  \\
		& Reg + no constraint & 0.29	 &	0.43	 &	0.15 &	0.14 &	5985 &	0.441  \\
		& With constraint & 0.29 &	0.45	 &	0.15 &	0.12 & 534 & 0.041  \\		
		\hline
	\end{tabular}
	
\medskip
The `initial conditions' row gives the fractional number count orbit classification while the other rows giving the orbit weight classification. \rjlr{Red text} indicates differences of 3\% or more between model values (with or without regularization) and the \cite{Xu2019} values.  In this case, 53\% (21/40) of the values compared differ.  In general, the models do not reproduce consistently the target orbit classifications from \cite{Xu2019}.  Also, the regularized and unregularized models themselves do not agree on their orbit classifications for a galaxy.  The `with constraint' row gives the orbit classification if the \cite{Xu2019} classification is used as a model constraint.  In all cases the model classification agrees with the \cite{Xu2019} classification with slight variations to the active orbit count and the mean $\chi^2$ values.
\end{table}
Looking at the mean $\chi^2$ column, the observed data values are reproduced well by the modeling.  For models without regularization, mean $\chi^2$ values $<< 1$ indicate that significant overfitting is taking place.  The heaviest weighted orbits for a galaxy are those orbits remaining close (in projection) to the galaxy center with the lighter weighted orbits having larger apocenters.  Figure \ref{fig:kins1centre} shows this weight progression for galaxy A1190 as an example. The position of the heaviest weights centrally matches the region where the surface brightness observable dominates numerically other observables in the model. 
Using a simple circularity classification into hot, warm and cold orbits with no separation by sense of rotation, it is clear that orbits in different classes tend to have a role to play in particular regions---see Figure \ref{fig:kins1A1390}, this time using galaxy A1390 as an example.

\begin{figure}[h]
\centering
\caption{Impact of NNLS Regularization and Smoothing}
\begin{subfigure}{0.48\textwidth}
    \includegraphics[width=\textwidth]{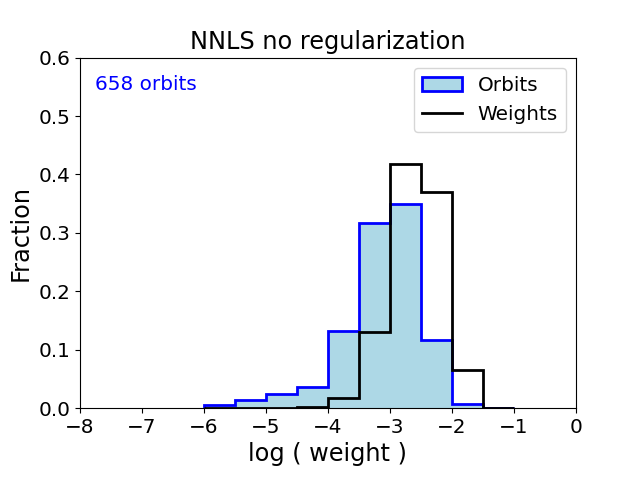}
\end{subfigure}
\begin{subfigure}{0.48\textwidth}
    \includegraphics[width=\textwidth]{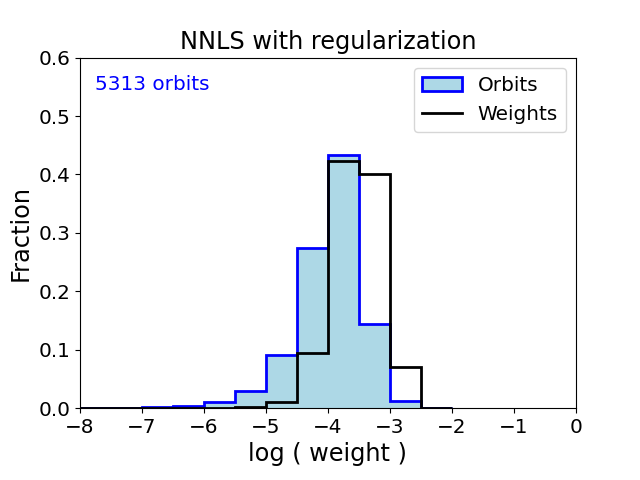}
\end{subfigure}
\begin{subfigure}{0.48\textwidth}
    \includegraphics[width=\textwidth]{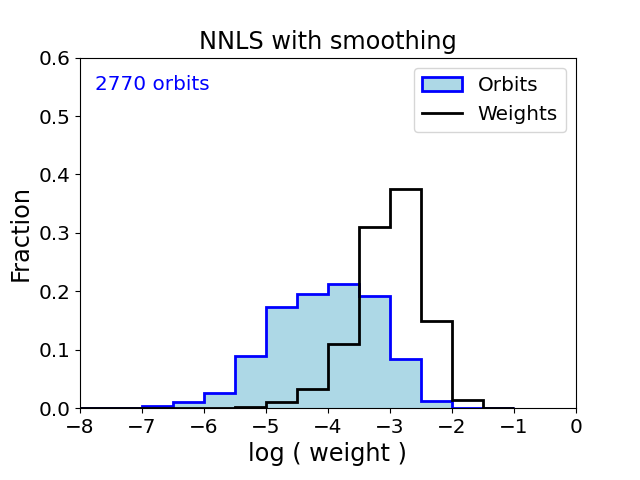}
\end{subfigure}
\begin{subfigure}{0.48\textwidth}
    \includegraphics[width=\textwidth]{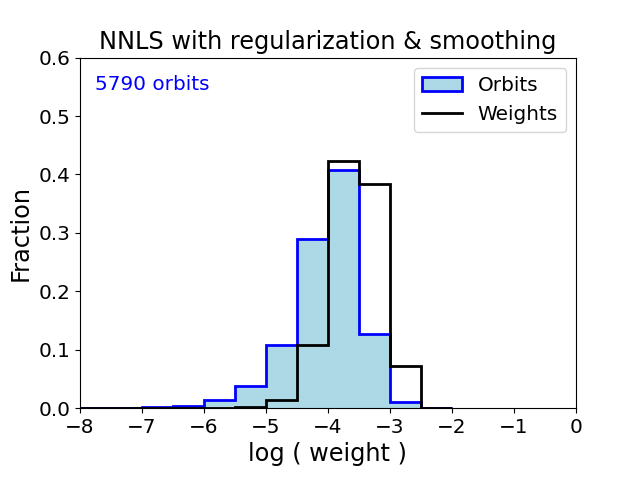}
\end{subfigure}

\medskip
All the plots are for simulated galaxy A1090.  The plots show how the orbit number and weight distributions change as regularization and smoothing are applied. See Section \ref{sec:cbaseline} for more information.
\label{fig:kins1reg}
\end{figure}

As regularization is penalizing heavy weights, the maximum orbit weights are reduced (by approximately an order of magnitude), and the weight ranges are extended to lower values.  Regularization does cause more orbits to be active.  However the issue concerning a small number of orbits being allocated heavy weights remains: it has just been `down-shifted' by a factor of 10---see the top row of Figure \ref{fig:kins1reg}.  
Smoothing alone also causes more orbits to be active but some orbits have even heavier weights as the orbit count distribution is broadened overall, generally to lower weights---see the bottom row of Figure \ref{fig:kins1reg}. Combining regularization and smoothing together produces a result more akin to regularization alone but with more active orbits, and a performance penalty (see Table \ref{tab:simoptelap}).

Overall, for the simulated galaxies, the models do not reproduce consistently the target orbit classifications from \cite{Xu2019}.  Also, the regularized and unregularized models themselves do not agree on their orbit classifications for a galaxy.  In all cases, if the \cite{Xu2019} classification is used as a model observable,  the model is able to reproduce that observable demonstrating that the classification is not unachievable in a modeling sense.  There is a minor difference regarding the determination of    orbit classifications between this work and \cite{Xu2019}: this work uses data extending to three effective radii while \cite{Xu2019} uses two effective radii.  This difference has been assessed as not significant.

The results for the \atlas\ galaxies are in Table \ref{tab:a3dgalres1}, and are consistent with the simulated galaxy results except that there are no target orbit classifications for comparison.
\begin{table}
	\centering
	\caption{Results for \atlas\ Galaxies}
	\label{tab:a3dgalres1}
	\begin{tabular}{cccccccc}
		\hline
		Galaxy & & Hot & Warm & Cold & Counter  & Active & Mean\\
		       & &     &      &      & Rotating & Orbits& $\chi^2$\\
		\hline
		NGC 1248 & \\			
		& Initial conditions & 0.19 &	0.41 &	0.20 &	0.20 &	8000  \\							
		& Without reg & 0.22	 & 	0.35 & 	0.34 & 	0.09 &	495 &	0.019  \\				
		& With reg & \rjlr{0.17} & 	\rjlr{0.41} & 	\rjlr{0.26} &	\rjlr{0.16} &	7493 & 	0.220  \\	
		NGC 3838 & \\					
		& Initial conditions & 0.19 &	0.41 &	0.20 &	0.20 &	8000	  \\					
		& Without reg & 0.18 &	0.35 &	0.30 &	0.16 &	679 &	0.020  \\					
		& With reg & 0.19 &	0.36 &	0.32 &	\rjlr{0.13} &	6342 &	0.248  \\	
		NGC 4452 & \\					
		& Initial conditions & 0.19 &	0.40 &	0.20	 &	0.20 &	8000  \\						
		& Without reg & 0.29 &	0.28 &	0.20 &	0.21 &	668 &	0.041  \\				
		& With reg & \rjlr{0.17} &	0.30 &	\rjlr{0.33} &	0.21 &	7191 &	0.250  \\
		NGC 4551 & \\				
		& Initial conditions & 0.19 &	0.40 &	0.21 &	0.20 &	8000  \\						
		& Without reg & 0.29 &	0.37 &	0.16 &	0.18 &	861 &	0.028  \\
		& With reg & \rjlr{0.24}	 &	0.37	 &	0.17 &	\rjlr{0.22} &	6618 &	0.195  \\	
		\hline
	\end{tabular}
	
\medskip
The description of the table rows and columns follows from Table \ref{tab:simgalres} except that there is no target orbit classification for comparison or constraint purposes.  \rjlr{Red text} indicates differences of 3\% or more between orbit classification values from models with and without regularization.  In this case, 56\% (9/16) of the values compared differ.  The number of active orbits and mean $\chi^2$ are as expected.
\end{table}

\subsection{Orbit Classifications from Grid-searching} 

When no target orbit classification is available, it is possible to construct a grid search of Schwarzschild models to determine an appropriate classification from the orbit data.  The search used here is to construct classifications by randomly uniformly selecting values from supplied minimum to maximum class ranges, and then to score those classifications using unregularized Schwarzschild modeling.  A number of classifications closest in (Euclidean) distance to the classification with the minimum score are used to set revised minimum to maximum class ranges, and the classification construction and scoring processes are repeated.  For the simulated galaxies, two iterations after the initial scoring are sufficient to have converged to an orbit classification for a galaxy.

The number of classifications created for scoring is 512, and 64 classifications are used to create the revised minimum to maximum class ranges.  The [hot, warm, cold, counter-rotating] class ranges used initially for the simulated galaxies are [(0.2, 0.5), (0.2, 0.5), (0.1, 0.25), (0.1, 0.25)].  For the \atlas\ galaxies, the range (0.1, 0.5) is used for all classes.  Scores are calculated as the total $\chi^2$ for the observables plus the orbit classification expression (\ref{eqn:extcirc1}).

The results achieved are shown in Table \ref{tab:galgrid}.
\begin{table}
	\centering
	\caption{Grid Results for Simulated and \atlas\ Galaxies}
	\label{tab:galgrid}
	\begin{tabular}{cccccc}
		\hline
		Galaxy & & Hot & Warm & Cold & Counter  \\
		       & &     &      &      & Rotating \\
		\hline
		A490 & Grid search & 0.37 &	0.29 &	0.15 &  	0.19  \\						
		& Without reg & 0.38 &	0.28 &	0.16 &	0.18  \\					
		A1090 & Grid search & 0.27 &	0.39 &	0.21 & 	0.13	  \\							
		& Without reg & 0.26	 &  0.39 & 	0.21 & 	0.14  \\				
		A1190 & Grid search & 0.39 &	0.34 &	0.14 &	0.13  \\					
		& Without reg & 0.39 &	0.34 &	0.14 &	0.13  \\					
		A1290 & Grid search & 0.27 &	0.34	 &	0.25 &	0.14  \\						
		& Without reg & 0.27 &	0.33 &	0.26 &	0.14  \\				
		A1390 & Grid search & 0.27 &	0.41 &	0.18 &	0.14  \\						
		& Without reg & 0.27 &	0.41 &	0.18 &	0.14  \\
		\hline
		NGC 1248 & Grid search & 0.25 &	0.30 &	0.29 &	0.17  \\							
		& Without reg & 0.25	 & 	0.29 & 	0.32 & 	0.14  \\				
		NGC 3838 & Grid search & 0.15 &	0.41 &	0.31 &	0.13  \\					
		& Without reg & 0.16 &	0.40 &	0.34 &	0.11  \\					
		NGC 4452 & Grid search & 0.28 &	0.31 &	0.25 &	0.17  \\						
		& Without reg & 0.27 &	0.31 &	0.26 &	0.16  \\				
		NGC 4551 & Grid search & 0.30 &	0.33 &	0.15 &	0.23  \\						
		& Without reg & 0.30 &	0.32 &	0.16 &	0.21  \\
		\hline
	\end{tabular}
	
\medskip
Row `grid search' indicates the orbit classifications determined using the grid search approach.  Row `without reg' repeats the orbit classifications from Tables \ref{tab:simgalres} and \ref{tab:a3dgalres2}.  For the simulated galaxies, the orbit classifications differ by at most $0.01$.  Some of the differences for the \atlas\ galaxies are $>0.01$, perhaps indicating that a 3rd search iteration is needed due to the wider cold and counter-rotating ranges.
\end{table}
The search classifications match the classifications from the baseline unregularized modeling runs in Tables \ref{tab:simgalres} and \ref{tab:a3dgalres2} quite well.  This is to be expected from numerical optimization considerations.  Pair-wise `corner' plots show a negative correlation between classification class values.  This is not unexpected as the sum of the weights in the classes
must be equal to $1.00$, so an increase in one class must be compensated for in the other classes.

\subsection{Initial Conditions---Rotation Extremes}
\label{sec:norot}
In Section \ref{sec:cbaseline}, all the models have initial conditions where there is a sense of bulk rotation aligned with the mean velocity data. In Table \ref{tab:a3dgalres2}, a set of results for \atlas\ galaxy models is shown where the models are lacking an overall sense of rotation. The actual \atlas\ galaxies do have a sense of rotation---see Table \ref{tab:atlasdata}.  
\begin{table}
	\centering
	\caption{Results for \atlas\ Galaxies---No Overall Sense of Rotation}
	\label{tab:a3dgalres2}
	\begin{tabular}{cccccccc}
		\hline
		Galaxy & & Hot & Warm & Cold & Counter  & Active & Mean\\
		       & &     &      &      & Rotating & Orbits& $\chi^2$\\
		\hline
		NGC 1248 & \\			
		& Initial conditions & 0.20 &	0.27 &	0.13 &	0.40 &	8000  \\							
		& Without reg & 0.25	 & 	0.29 & 	0.32 & 	0.14 &	451 &	0.022  \\				
		& With reg & \rjlr{0.18} & 	\rjlr{0.36} & 	\rjlr{0.23} &	\rjlr{0.23} &	7263 & 	0.251  \\	
		NGC 3838 & \\					
		& Initial conditions & 0.20 &	0.27 &	0.13 &	0.40 &	8000	  \\					
		& Without reg & 0.16 &	0.40 &	0.34 &	0.11 &	671 &	0.023  \\					
		& With reg & \rjlr{0.19} &	\rjlr{0.34} &	\rjlr{0.30} &	\rjlr{0.16} &	5712 &	0.273  \\	
		NGC 4452 & \\					
		& Initial conditions & 0.19 &	0.27 &	0.14	 &	0.40 &	8000  \\						
		& Without reg & 0.27 &	0.31 &	0.26 &	0.16 &	642 &	0.045  \\				
		& With reg & \rjlr{0.17} &	\rjlr{0.26} &	\rjlr{0.31} &	\rjlr{0.26} &	7048 &	0.263  \\
		NGC 4551 & \\				
		& Initial conditions & 0.19 &	0.28 &	0.13 &	0.40 &	8000  \\						
		& Without reg & 0.30 &	0.32 &	0.16 &	0.21 &	856 &	0.026  \\
		& With reg & \rjlr{0.24}	 &	0.34	 &	0.15 &	\rjlr{0.27} &	6598 &	0.195  \\	
		\hline
	\end{tabular}
	
\medskip
The description of the table rows and columns follows from Table \ref{tab:simgalres} except that there is no target orbit classification for comparison or constraint purposes.  \rjlr{Red text} indicates differences of 3\% or more between orbit classification values from models with and without regularization.  In this case, 88\% (14/16) of the values compared differ.  The number of active orbits and mean $\chi^2$ are as expected.
\end{table}
No sense of rotation implies that 50\% of the orbits are rotating in the same sense as the velocity data and 50\% are counter-rotating with the opposite sense.   Within the table, counter-rotating orbits are split between the counter-rotation class (40\%) and the hot class (10\%).  All the models appear to be acceptable (mean $\chi^2 < 1$).  The orbit classifications between Tables \ref{tab:a3dgalres1} and \ref{tab:a3dgalres2} do not agree, however.  Our assessment is that, while it may be acceptable to allow the modeling method to decide what rotation mix is satisfactory for determining global galaxy attributes, for orbit classification investigations it is better to explicitly determine the sense of rotation which matches the observed data.

The fully aligned scheme, where all orbits are aligned with a galaxy's sense of rotation, does not perform well: all the simulated galaxy models fail to reproduce the observed data satisfactorily.  For the five galaxies [A490, A1090, A1190, A1290, A1390], using regularized modeling, the mean model $\chi^2$ values are [$4.59$, $1.55$, $1.66$, $1.58$, $1.50$] with all galaxies having their maximum individual observable mean $\chi^2$ values greater than $2.75$, and all galaxies having at least three individual observable mean $\chi^2$ values $>1.5$.  Given the high mean $\chi^2$ values, the model orbit classifications are not displayed here.

In summary, where the kinematical data exhibit bulk rotation of a galaxy, fully aligned rotation is unlikely to be satisfactory: some counter-rotating orbits appear to be necessary to reproduce the observed data.  No sense of rotation may be appropriate if no overall rotation or only very minor rotation is observed.  This leaves modeling of many galaxies in a position where some overall sense of rotation should be applied to the initial conditions but just how much ought to be determined via experimentation.

\subsection{Initial Conditions---An Alternative to 3I}
An alternative scheme for initial conditions is the MDJV, match density Jeans velocities, scheme described in \cite{Long2021}.  The motivation for MDJV is to use the observed data in setting the initial conditions.  A set of models of the simulated galaxies using MDJV is constructed and compared with the equivalent 3I baseline models in Section \ref{sec:cbaseline}.

\begin{table}
	\centering
	\caption{Results for Simulated Galaxies using MDJV}
	\label{tab:simgalmjdv}
	\begin{tabular}{cccccccc}
		\hline
		Galaxy & & Hot & Warm & Cold & Counter  & Active & Mean\\
		       & &     &      &      & Rotating & Orbits & $\chi^2$\\
		\hline
		A490 &  \\					
		& Initial conditions & 0.27 &	0.45 &	0.09 &  	0.18 &	8000  \\	
		& \cite{Xu2019} & 0.35 &	0.28 &	0.14 &  	0.23  \\						
		& Without reg & \rjlr{0.42} &	0.30 &	\rjlr{0.10} &	\rjlr{0.18} &	797 &	0.046  \\					
		& With reg & \rjlr{0.38}	 &  \rjlr{0.34}	 &	\rjlr{0.09} & 	\rjlr{0.20} &	6078 & 	0.277  \\
		A1090 & \\			
		& Initial conditions & 0.32 &	0.45 &	0.06 &	0.17 &	8000  \\	
		& \cite{Xu2019} & 0.28 &	0.46 &	0.15 & 	0.10	  \\							
		& Without reg & \rjlr{0.32}	 & 	\rjlr{0.43} & 	0.13 & 	0.12 &	742 &	0.033  \\				
		& With reg & \rjlr{0.34} & 	0.46 & 	\rjlr{0.09} &	0.11 &	6831 & 	0.292  \\	
		A1190 & \\					
		& Initial conditions & 0.36 &	0.44 &	0.04 &	0.16 &	8000	  \\	
		& \cite{Xu2019} & 0.33 &	0.43 &	0.11 &	0.12  \\					
		& Without reg & \rjlr{0.47} &	\rjlr{0.37} &	\rjlr{0.06} &	0.11 &	634 &	0.004  \\					
		& With reg & \rjlr{0.43} &	0.41 &	\rjlr{0.03} &	0.13 &	7330 &	0.205  \\		
		A1290 & \\					
		& Initial conditions & 0.28 &	0.43 &	0.10	 &	0.18 &	8000  \\	
		& \cite{Xu2019} & 0.32 &	0.37	 &	0.18 &	0.13  \\						
		& Without reg & 0.34 &	0.36 &	0.17 &	0.13 &	777 &	0.011  \\				
		& With reg & 0.34 &	0.38 &	\rjlr{0.15} &	0.13 &	7052 &	0.229  \\	
		A1390 & \\				
		& Initial conditions & 0.33 &	0.44 &	0.06 &	0.17 &	8000  \\	
		& \cite{Xu2019} & 0.29 &	0.45 &	0.15 &	0.12  \\						
		& Without reg & \rjlr{0.36} &	\rjlr{0.42} &	\rjlr{0.10} &	0.11 &	597 &	0.015  \\
		& With reg & \rjlr{0.36}	 &	0.46	 &	\rjlr{0.06} &	0.12 &	7271 &	0.279  \\	
		\hline
	\end{tabular}
	
\medskip
The description of the table rows and columns follows from Table \ref{tab:simgalres}.  \rjlr{Red text} indicates differences of 3\% or more between model values and the \cite{Xu2019} values.  In this case, 55\% (22/40) of the values compared differ.  In general, the models, with or without regularization, do not reproduce the target orbit classifications from \cite{Xu2019}.  Also, the regularized and unregularized models do not agree on the orbit classification for a galaxy.
\end{table}

Comparing the MDJV results in Table \ref{tab:simgalmjdv} with the 3I results in Table \ref{tab:simgalres}, the mean model $\chi^2$ values for the MDJV models are all lower than their 3I counterparts indicating that overfitting has increased.  The maximum mean $\chi^2$ value for MDJV is 0.29 while the maximum for 3I is 0.64.  The number of active orbits is increased with the MDJV models.
From the results achieved, there is no reason to doubt the validity of using MDJV initial conditions.  MDJV does produce orbit classifications that do not agree with those produced by the 3I scheme.

Note that the orbit count circularity classifications for the two schemes are different with the MDJV models having fewer cold orbits than the 3I models. For the five galaxies [A490, A1090, A1190, A1290, A1390], the cold orbit number percentages are [12.3\%, 8.2\%, 6.0\%, 14.3\%, 8.4\%] for MDJV and [20.7\%, 19.9\%, 20.1\%, 19.4\%, 19.4\%] for 3I.  It is not obvious that this number difference directly influences the model orbit classification.

\subsection{Removal of Orbit Classes}
In this section, all orbits of a given orbit circularity class are removed from an existing model and the weighting process rerun.  The resulting model is then re-assessed for the impact on reproduction of observables only.  Three sets of models are created in this way - the removal of all the cold orbits from the \atlas\ models in Section \ref{sec:norot}, cold orbit removal from the simulated galaxy baseline models in Section \ref{sec:cbaseline}, and the removal of hot orbits from the simulated galaxy baseline models.  To be clear, removing cold orbits means modeling with hot and warm orbits only.  Similarly, removing hot orbits means modeling with warm and cold orbits only.

With the removal of cold orbits from the baseline models, for the NNLS regularized models of galaxies [A490, A1090, A1190, A1290, A1390], the mean model $\chi^2$ values are $[0.99, 0.83, 0.47, 1.50, 0.68]$.  Looking at the individual observable mean $\chi^2$ values, the observables for galaxy A1290 have not been well reproduced (one observable $\chi^2$ value $>2$, four $>1$).  Reproduction of observables for the other four galaxies is satisfactory.
Removing hot orbits from the baseline models, the mean model $\chi^2$ values are $[1.47, 0.78, 1.06, 091, 0.85]$.  Observables have not been well reproduced for galaxies A490 and A1190; for the other three galaxies, reproduction is satisfactory.

With the removal of cold orbits from the \atlas\ models for galaxies [NGC 1248, NGC 3838, NGC 4452, NGC 4551], the mean model $\chi^2$ values are $[0.34, 0.71, 1.39, 0.21]$.  Observable reproduction for NGC 4452 is close to unacceptable (all observable mean $\chi^2$ are $>1$); there are no concerns for the other three galaxy models.

The models here show a strength of Schwarzschild's method in that it will attempt to weight the orbits it is given to try and reproduce observational data.  Whether the orbits are astrophysically realistic, either individually or collectively, is not its concern.

\subsection{Computer Utilization}
\label{sec:computil}

The software base for constructing initial conditions and Schwarzschild's modeling is taken from the lead author's implementation of the \cite{Syer1996} made-to-measure stellar dynamical modeling method and the \cite{Schwarz1979} orbit based modeling method.  This software was first used in \cite{Long2010}, and more recently in \cite{Long2021}.  Additional analysis code specific to this investigation is written in Python 3 \citep{10.5555/1593511}.  The optimization software used for NNLS is taken from the SCIPY package \citep{2020SciPy-NMeth}.  The functions used are untailored with default parameter values being taken.

Modeling runs were performed on a $20$ core desktop PC.  No attempt was made to use graphics processing units (GPUs) to improve performance but their use is not precluded.  All software used is Python 3 based with some use of Cython \citep{behnel2010cython} for performance critical code. Multi-processor working, limited to a maximum of 10 processors, was employed.  The elapsed times for orbit integration and data collection, and for weight determination (optimization) of the simulated galaxies are shown in Table \ref{tab:simoptelap} and in Table \ref{tab:atoptelap} for the \atlas\ galaxies.  The elapsed times correspond to stages one and two in Section \ref{sec:approach}.

\begin{table}
	\centering
	\caption{Orbit Integration and Weight Determination Elapsed Times for Simulated Galaxies}
	\label{tab:simoptelap}
	\begin{tabular}{|c|rr|rr|rr|rr|rr|}
		\hline
		\multicolumn{1}{|c}{} &		
		\multicolumn{2}{|c|}{\textbf{A490}} & \multicolumn{2}{c|}{\textbf{A1090}} & \multicolumn{2}{c|}{\textbf{A1190}} & 
		\multicolumn{2}{c|}{\textbf{A1290}} & \multicolumn{2}{c|}{\textbf{A1390}}\\
		\hline
		\multicolumn{1}{|l|}{\textbf{Integration}}  & 536 s & & 564 s & & 659 s & & 522 s & & 590 s &\\	
		\hline
		\multicolumn{1}{|l|}{\textbf{Weights - NNLS}} & & & & & & & & & &\\
		Unreg &  20 s & 9.3\%   & 15 s  & 8.2\%   & 11 s  & 7.0\%   & 17 s  & 8.4\%   & 11 s  & 6.6\%\\
		Heavy weights & 435 s & 59.0\%  & 448 s & 66.4\%  & 441 s & 73.0\%  & 437 s & 62.9\%  & 447 s & 74.8\%\\
		Smoothed & 731 s & 32.3\%  & 775 s & 34.6\%  & 792 s & 34.7\%  & 780 s & 33.5\%  & 807 s & 34.3\%\\
		Both & 1450 s & 62.4\% & 1532 s & 72.4\% & 1545 s & 76.3\% & 1472 s & 66.5\% & 1563 s & 78.7\%\\
		\hline
	\end{tabular}
	
\medskip
Elapsed times are shown in seconds.  The percentages indicate the percentage of active orbits (that is the orbits allocated non-zero weights) out of a total of 8000.  Orbits are integrated for 50 half-mass dynamical time units.  Within the table `Both' means that both heavy weight penalization and  integral space smoothing have been applied.  NNLS times depend on the regularization in use and the number of active orbits needed by the NNLS algorithm for reproducing observables.
\end{table}

\begin{table}
	\centering
	\caption{Orbit Integration and Weight Determination Elapsed Times for \atlas\ Galaxies}
	\label{tab:atoptelap}
	\begin{tabular}{|c|rr|rr|rr|rr|}
		\hline
		\multicolumn{1}{|c}{} &		
		\multicolumn{2}{|c|}{\textbf{NGC 1248}} & \multicolumn{2}{c|}{\textbf{NGC 3838}} & \multicolumn{2}{c|}{\textbf{NGC 4452}} & 
		\multicolumn{2}{c|}{\textbf{NGC 4551}}\\
		\hline
		\multicolumn{1}{|l|}{\textbf{Integration}}  & 321 s & & 481 s & & 385 s & & 188 s &\\	
		\hline
		\multicolumn{1}{|l|}{\textbf{Weights - NNLS}} & & & & & & & & \\
		Unreg &  11 s & 5.6\%   & 22 s  & 8.4\%   & 20 s  & 8.0\%   & 24 s  & 10.7\% \\
		Heavy weights & 478 s & 90.8\%  & 457 s & 71.5\%  & 501 s & 88.1\%  & 501 s & 82.6\% \\
		\hline
	\end{tabular}
	
\medskip
The table layout is the same as for Table \ref{tab:simoptelap} but without the weight smoothing models (which were not created for the \atlas\ galaxies). Orbits are integrated for 100 half-mass dynamical time units. 
\end{table}

\section{Discussion}
\label{sec:discussion}
As set out in the Introduction, Section \ref{sec:intro}, the science matter is concerned with how to gain
knowledge of the internal stellar orbital structure of external galaxies in
spite of the inability to obtain 6D phase space data, and to observe stellar orbits.
Astronomers have only blended observational data to rely on to try and gain knowledge
of structures that cannot be directly validated.

One solution is to try and use an approach such as Schwarzschild's method, and to assume
that the orbits that are selected give a good realistic representation of a galaxy, but 
this is based on assuming that the gravitational potential and the pool of orbits to be
selected from have the capacity to represent the galaxy.  At best, any such representation
can only be indicative or illustrative of the galaxy, and may be significantly inaccurate.

The advent of large language models (LLMs), a type of generative artificial intelligence
application, introduces another complication.  Astronomers have to be very careful about 
caveating their results appropriately and not, for example, overstating their results.  
Astronomy articles form part of the input to LLMs, and new researchers already risk being 
misinformed by unwisely formulated science articles.

Schwarzschild's method is good at trying to reproduce observational data, but this does 
not mean that the orbits selected to do so can be used for other purposes, unless the
selection process has been constructed appropriately.  This article offers a partial way forward
with orbit classifications being included in the selection process.  However, the results
here also show that it is clear that Schwarzschild's method can not be the source of those
classifications.  Just changing the orbit initial conditions, or using or not using regularization,
for example, is sufficient to change the orbit classification for a galaxy.

Using a machine learning application may be an option, if there is some data source which can be used 
to train the application.  Perhaps cosmological simulations will in time become that source.
Alternatively, perhaps the reasons why the internal stellar orbital structure is interesting 
should be re-examined.  Where orbital structure is a stepping stone, could it be skipped, for example?

\section{Conclusions}
\label{sec:conclusions}

The objective set out in the Introduction, Section \ref{sec:intro}, has been met. The question posed was `could an orbit classification for a galaxy being investigated be included in Schwarzschild modeling as a constraint?'  As has been demonstrated,
the answer is 'yes.'  Unfortunately, quite where that orbit classification might be obtained from is an open question: it does not appear to be from using Schwarzschild's method given its usual implementation; it might be from cosmological simulations.   Having the capability to include classification constraints is, however, another useful technique in the stellar dynamical toolbox.

\begin{acknowledgements}
The author thanks Dandan Xu, Ling Zhu and Shude Mao for various fruitful discussions as the research recorded here progressed.
This work is partly supported by the National Key Basic Research and Development Program of China (No. 2018YFA0404501 to Shude Mao), and by the National Natural Science Foundation of China (NSFC, grant Nos. 11821303, 11761131004 and 11761141012 to Shude Mao).  
\end{acknowledgements}

\bibliographystyle{raa}
\bibliography{ms2024-0327}

\begin{thebibliography}{34}
\providecommand\natexlab[1]{#1}
\providecommand\JournalTitle[1]{#1}

\bibitem[Behnel {et~al.}(2011)]{behnel2010cython}
Behnel, S., Bradshaw, R., Citro, C., {et~al.} 2011, Computing in Science
  Engineering, 13, 31

\bibitem[{Binney} \& {Tremaine}(2008)]{BT2008}
{Binney}, J., \& {Tremaine}, S. 2008, {Galactic Dynamics: Second Edition}
  (Princeton University Press)

\bibitem[{Boyd} \& {Vandenberghe}(2004)]{CVX}
{Boyd}, S., \& {Vandenberghe}, L. 2004, {Convex Optimization} (Cambridge
  University Press)

\bibitem[{Cappellari}(2002)]{Cappellari2002}
{Cappellari}, M. 2002, \mnras, 333, 400

\bibitem[{Cappellari}(2020)]{Cappellari2020}
{Cappellari}, M. 2020, \mnras, 494, 4819

\bibitem[{Cappellari} {et~al.}(2011)]{AtlasI}
{Cappellari}, M., {Emsellem}, E., {Krajnovi{\'c}}, D., {et~al.} 2011, MNRAS,
  413, 813

\bibitem[{Cappellari} {et~al.}(2013)]{AtlasXV}
{Cappellari}, M., {Scott}, N., {Alatalo}, K., {et~al.} 2013, MNRAS, 432, 1709

\bibitem[{Chanam{\'e}} {et~al.}(2008)]{Chaname2008}
{Chanam{\'e}}, J., {Kleyna}, J., \& {van der Marel}, R. 2008, \apj, 682, 841

\bibitem[{Cretton} {et~al.}(1999)]{Cretton1999}
{Cretton}, N., {de Zeeuw}, P.~T., {van der Marel}, R.~P., \& {Rix}, H.-W. 1999,
  \apjs, 124, 383

\bibitem[{Emsellem} {et~al.}(1994)]{Emsellem1994}
{Emsellem}, E., {Monnet}, G., \& {Bacon}, R. 1994, A\&A, 285, 723

\bibitem[{Jin} {et~al.}(2020)]{Jin2020}
{Jin}, Y., {Zhu}, L., {Long}, R.~J., {et~al.} 2020, \mnras, 491, 1690

\bibitem[{Jin} {et~al.}(2019)]{Jin2019}
{Jin}, Y., {Zhu}, L., {Long}, R.~J., {et~al.} 2019, \mnras, 486, 4753

\bibitem[{Li} {et~al.}(2016)]{Li2016}
{Li}, H., {Li}, R., {Mao}, S., {et~al.} 2016, \mnras, 455, 3680

\bibitem[{Long}(2016)]{Long2016}
{Long}, R.~J. 2016, Research in Astronomy and Astrophysics, 16, 189

\bibitem[{Long} {et~al.}(2021)]{Long2021}
{Long}, R.~J., {Lu}, S.-D., \& {Xu}, D.-D. 2021, Research in Astronomy and
  Astrophysics, 21, 152

\bibitem[{Long} \& {Mao}(2010)]{Long2010}
{Long}, R.~J., \& {Mao}, S. 2010, MNRAS, 405, 301

\bibitem[{Long} \& {Mao}(2012)]{Long2012}
{Long}, R.~J., \& {Mao}, S. 2012, MNRAS, 421, 2580

\bibitem[{Long} \& {Mao}(2018)]{Long2018}
{Long}, R.~J., \& {Mao}, S. 2018, Research in Astronomy and Astrophysics, 18,
  145

\bibitem[{Nelson} {et~al.}(2019)]{Nelson2019}
{Nelson}, D., {Springel}, V., {Pillepich}, A., {et~al.} 2019, Computational
  Astrophysics and Cosmology, 6, 2

\bibitem[Press {et~al.}(2007)]{NumRecipes2007}
Press, W.~H., Teukolsky, S.~A., Vetterling, W.~T., \& Flannery, B.~P. 2007,
  Numerical Recipes 3rd Edition: The Art of Scientific Computing, 3rd edn.
  (USA: Cambridge University Press)

\bibitem[{Raissi} {et~al.}(2019)]{Raissi2019}
{Raissi}, M., {Perdikaris}, P., \& {Karniadakis}, G.~E. 2019, Journal of
  Computational Physics, 378, 686

\bibitem[{Rix} {et~al.}(1997)]{Rix1997}
{Rix}, H.-W., {de Zeeuw}, P.~T., {Cretton}, N., {van der Marel}, R.~P., \&
  {Carollo}, C.~M. 1997, ApJ, 488, 702

\bibitem[{Rybicki}(1987)]{Rybicki1987}
{Rybicki}, G.~B. 1987, in IAU Symposium, Vol. 127, Structure and Dynamics of
  Elliptical Galaxies, ed. P.~T. {de Zeeuw}, 397

\bibitem[{Schwarzschild}(1979)]{Schwarz1979}
{Schwarzschild}, M. 1979, ApJ, 232, 236

\bibitem[{Syer} \& {Tremaine}(1996)]{Syer1996}
{Syer}, D., \& {Tremaine}, S. 1996, MNRAS, 282, 223

\bibitem[{Tikhonov}(1963)]{Tikhonov1963}
{Tikhonov}, A.~N. 1963, Soviet Mathematics, 4, 1035

\bibitem[{Valluri} {et~al.}(2004)]{Valluri2004}
{Valluri}, M., {Merritt}, D., \& {Emsellem}, E. 2004, \apj, 602, 66

\bibitem[{van den Bosch} {et~al.}(2008)]{Remco2008}
{van den Bosch}, R.~C.~E., {van de Ven}, G., {Verolme}, E.~K., {Cappellari},
  M., \& {de Zeeuw}, P.~T. 2008, MNRAS, 385, 647

\bibitem[Van~Rossum \& Drake(2009)]{10.5555/1593511}
Van~Rossum, G., \& Drake, F.~L. 2009, Python 3 Reference Manual (Scotts Valley,
  CA: CreateSpace)

\bibitem[{Vasiliev}(2013)]{Vasiliev2013}
{Vasiliev}, E. 2013, MNRAS, 434, 3174

\bibitem[{Vasiliev} \& {Valluri}(2020)]{Vasiliev2020}
{Vasiliev}, E., \& {Valluri}, M. 2020, \apj, 889, 39

\bibitem[Virtanen {et~al.}(2020)]{2020SciPy-NMeth}
Virtanen, P., Gommers, R., Oliphant, T.~E., {et~al.} 2020, Nature Methods, 17,
  261

\bibitem[{Xu} {et~al.}(2019)]{Xu2019}
{Xu}, D., {Zhu}, L., {Grand}, R., {et~al.} 2019, \mnras, 489, 842

\bibitem[{Zhu} {et~al.}(2018)]{Zhu2018}
{Zhu}, L., {van de Ven}, G., {van den Bosch}, R., {et~al.} 2018, Nature
  Astronomy, 2, 233

\end{thebibliography}

\end{document}